# Why ferromagnetic semiconductors?[*]

Tomasz Dietl[**]

*Institute of Physics, Polish Academy of Sciences,
al. Lotników 32/46, PL-02-668 Warszawa, Poland*

**Abstract:** Rapid development of information technologies originates from the exponential increase in the density of information that can be processed, stored, and transfer by the unit area of relevant devices. There is, however, a growing amount of evidences that the progress achieved in this way approaches its limits. Various novel ideas put forward to circumvent barriers ahead are described. Particular attention is paid to those concepts which propose to exploit electron or nuclear spins as the information carriers. Here, ferromagnetic semiconductors of III-V or II-VI compounds containing a sizable concentration of transition metals appear as outstanding spintronic materials.

________

## 1. Introduction

The information revolution, which has surprised us over the last few decades, has occurred due to the enormous progress in means of information processing, storing, and transmitting. Thus, it is not astonishing that the Nobel Prize Committee decided to single out in 2000 milestone concepts of Zores I. Alferov, Jack S. Kilby, and Herbert Kroemer. Actually, the outcome of the laureate ideas is surpassing today the speculations of Richard Feynman about manipulating and controlling things on a small scale [1], regarded as totally unrealistic in time of his visionary speech 40 years ago. Kilby [2] was among those who put forward the notion of integrated circuits. By now, semiconductor processors and memories, despite containing several hundreds of millions of submicron elements, are both astonishingly cheap and redundant. Alferov [3] and Kroemer [4] noted, in turn, that structures consisting of layers of various semiconductors make it possible to obtain faster transistors and more efficient lasers than those containing only homojunctions. Heterostructures constitute today basic building blocks of emitters and receivers in cellular, satellite, and fibreglass communication. It is thus obvious to regard the 2000 Nobel Prize as honouring the achievements of the information age but, perhaps, also as a closure of the passing epoch.

Indeed, there is a growing amount of evidences that the progress achieved by miniaturization of transistors and memory cells is starting to exhaust its possibilities.

The first part of this review, based on earlier author's papers [5,6], is devoted to the description of barriers that may slow down the exponential growth in efficiency of information devices persisting since 60's. Various suggestions concerning the future of nanotechnology are then discussed [6]. Particular attention is paid to spin electronics (spintronics) that aims at developing devices in which the electron charge and spin are exploited on the same footing. After enlisting spintronic goals, the most promising material families are described. Particularly important, from the point of view of both basic and applied science, appear to be ferromagnetic semiconductors, whose outstanding properties and expected functionalities are presented in the last chapter of the paper.

## 2. Prospects of nanotechology

Despite the fact that – as someone noted – predictions are not difficult unless concern the future, it is tempting to contemplate various scenarios according to which information technologies (IT) may develop. One is that, which occurred in the case of transoceanic travels – as it is known, during the last 20 years or so the speed of civil jets has *not* changed. Also in the case of IT, the financial barriers (a new chip plant is now a multibillion investment), the psychological barriers (fear of new; no real needs for further improvements), the legal barriers (intellectual property rights; dissemination of terrorism and pornography), …, might lead to the break down of Moor's low, according to which the number of elements per unit area of the device doubles each 18 months.

There is a consensus, however, that needs known and unknown today will drive the progress in IT for a long time ahead. For instance, it appears that the shift from the present gigaflop computations (one billion of operations per second) to pentaflop computations will open new prospects in the field of both virtual entertainment and simulations of physical, geophysical, chemical, biological, and social systems as well as – due to the improved pattern recognition – in the vehicle routing. It is worth recalling at this point that although the computational achievements appears as impressive indeed, we are now able to simulate from first principles the evolution of systems containing only of a few hundreds of atoms over a time shorter than 10 ps.

Is thus further progress possible by continuing miniaturization of, say, field-effect silicon transistors (MOS FET) or ferromagnetic and optical memory cells? Some experts suggest that today's technology and its development according to Moore's low will last no longer than 5 to 10 years. This stems from a number of technical problems that have to be overcome in order to push further the performance of the devices. Some of obstacles are not very spectacular – for instance, it is easier now to accelerate the microprocessor than to reduce the delay time of signal transmission through its case. There is, of course, a spectrum of fundamental barriers – the diffraction limit in lithography, the grain nature of matter and charge, the quantum phenomena such as tunnelling of electrons through insulators, the thermodynamic effects leading to jumps of magnetization between two bit states of the ferromagnetic nanoparticle as well as the heat dissipation associated with the switching process of the transistor.

In this situation, a number of industrial, national, and university laboratories has undertaken research in the field of nanostructures or more generally in the field of nanotechnology or nanoscience. However, despite this concerted effort, it is still unclear what sort of technology will dominate in the future. It is only known that the integration of elements will be more and more accompanied by the integration of functions: processors with memories, sensors with actuators, electronic and photonic devices with magnetic and nonmechanical devices. At the same time, lithography will be replaced by synthesis of elements, and the planar technology by a three dimensional architecture. An increasing role will be played by self-organized growth in epitaxy as well as by organic and biology synthesis, supplemented by new methods of single atom and molecule manipulation. This will bring us closer to practical realization of idea of molecular electronics. In view of health care needs, particularly important will be the development of means enabling the integration of biological cells with electronic and mechanical nanodevices.

Substantial changes will not omit the information carrier. Today, the charge of the electron serves for information processing, while its internal momentum (spin) is used for information storing. It appears that the growing role will be played by the photons, which are already employed for information transmission, coding, and writing. In addition to the electrons and photons, a number of respected scientists associate much hope with current vortices in type II superconductors (Abrikosov vortices) and with magnetic field fluxes in superconducting nanosolenoids.

Looking on the future of nanotechnology from the material view-point, one sees a growing interest in heterostructures of silicon, germanium and carbon. Such a material system not only makes it possible to speed up the transistors, but might also extend the domination of elemental semiconductors towards photonics, where the key materials have been III-V compounds, such as GaAs. Because of temperature stability and light generation up to the ultraviolet spectral range, an important role will be played by SiC- and GaN-based compounds as well as by … diamond. With no doubt a growing importance is predicted for countless family of organic compounds, employed with such a success by brain. The discovery of high temperature superconductors has directed much attention towards oxides. These materials are not only superconducting but exhibit also outstanding magnetic characteristics, so that one hears often about oxide electronics.

Along with alterations of device operation principles that will certainly make use of phenomena regarded as unwanted today – tunnelling, interference, grain structure of charge – new breakthrough in the computer architecture are to be expected. It seems, in particular, that various phenomena in disordered, chaotic, chemical, and biological systems, which we simulate tediously by today's computers, will serve tomorrow for fast computations according to algorithms imposed by the character of the phenomenon. Especially attractive is another visionary Feynman's idea [7] about the computation – the quantum computation -- proceeding according to the time-dependent Schrödinger equation for a given system. It may also turn out that instead of rejecting chips with errors, the interconnections in integrated circuits will be created randomly, and the application of particular chips will be determined *a posteriori*.

### 3. Spintronics

Spin electronics (spintronics) is a young interdisciplinary field of nanoscience. Its rapid development, like that of competing new branches of electronics – molecular electronics, bioelectronics, and electronics of polymers, …, has its roots in the conviction that the progress that is being achieved by miniaturization of active elements (transistors and memory cells) cannot continue forever. Therefore, the invention of future IT must involve new ideas concerning the design of both devices and system architecture.

The main goal of spintronics is to gain knowledge on spin-dependent phenomena, and to exploit them for new functionalities. Hopes associated with spintronics stem from the well-known fact that the magnetic fields present in the ambient world are significantly weaker than the electric fields. For this reason magnetic memories are non-volatile, while memories based on the accumulated electric charge (Dynamic Random Access Memory - DRAM) require a frequent refreshing.

At present, one can already specify a set of problems to be solved by spin electronics. One of them is a construction of efficient micro-sensors of the magnetic field, which would replace devices employing magnetic coils. It is obvious that the increase in spatial resolution requires a reduction in size of the sensor. In the case of the coil, however, this is accompanied by a decrease in the sensitivity. Intensive research and development work carried out over last fifteen years or so, resulted in the developing of the appropriate device. The new-generation sensors are exploiting a giant magnetoresistance effect (GMR) of multiple layer structures made of alternating ferromagnetic, antiferromagnetic, and paramagnetic metals [8,9]. The GMR results from the increase of electrical conductivity in the presence of the magnetic field that aligns the direction of magnetization vectors in neighboring layers. In contrast to the traditional sensors containing the Hall probe, the operation of the GMR sensor depends on the electron spin, and not on the charge, *i.e.,* the Lorenz force.

The most recent research on the field sensors focuses on spin-dependent electron tunneling between ferromagnetic layers through an isolator, typically thin aluminum oxide. It is expected that the tunneling effect will lead to a significant increase of magnetoresistance [10,11]. A side issue concerns the phenomenon of the Coulomb blockade that plays an important role whenever a tunnel junction becomes small [11]. One can expect that the parameters of tunneling magnetoresistance (TMR) sensors will make them suitable not just for reading devices, but also as position detectors, for instance, in electric and gasoline engines, where Hall-effect sensors dominate today [12].

A much more ambitious spintronic objective is to develop magnetic random access memories (MRAM). These devices would combine the advantages of magnetic memories and those of DRAM. For this purpose, it is necessary to find means of writing and reading the direction of magnetization in given cells without employing any moving parts. An important step would be the elaboration of methods of controlling magnetization isothermally – by light or electric field, like in semiconductor DRAM, in which information writing proceeds *via* voltage biasing of the addressed transistor. In the magnetic memories presently available, the magnetization switching requires rather large power, as it is triggered either by the magnetic field generated by electric currents or by the laser heating above the Curie temperature. Obviously, research in this direction combines materials science, nanotechnology, physics of mesoscopic and strongly correlated systems.

Elaboration of „intelligent" methods enabling magnetization control would also make it possible to fabricate spin transistors [13,14]. This device consists of two ferromagnetic metals separated by a nonmagnetic conductor. Simple considerations demonstrate that if spin-polarized carriers injected to the nonmagnetic layer conserve spin orientation, the device resistance will depend on the relative directions of magnetization in the two ferromagnetic layers. Since the switching process does not involve any change in the carrier density, this transistor will be characterized by a favorable value of the product of electric power consumption and switching time, provided that the spin injection would be efficient and no mechanisms of spin relaxation would operate.

Perhaps the most important challenge for spintronics is the development of quantum computation and communication [15]. Particular importance of spin degree of freedom in this context originates from the fact that it can preserve phase coherence for a much longer time than the orbital degrees of freedom. Thus, the electron spin is much more promising than the electron charge for materialization of the present revolutionary ideas on quantum computing, quantum cryptography, data compression and teleportation [15]. Thus, spin quantum devices, such as quantum dots [16,17], may change not only the principles of devices operation, but also the basis of the half-century-old computer architecture. Some researchers suggest that the best candidate to carry quantum information would be nuclear spins of $^{31}P$ in isotopicly pure $^{28}Si$, where the spin coherence time reaches several hours [18]. However, it has also been shown experimentally that the spin polarization lifetime in doped semiconductors may be several orders of magnitude longer than the time of momentum relaxation [19]. One has to mention, although, that up to now the research on quantum computation is theoretical. Any experimental achievement, independently of the material used and the experimental method applied will be regarded as a significant breakthrough.

### 4. Ferromagnetic semiconductors

Today's research on spin electronics involves virtually all material families, the most mature being studies on magnetic metal multilayers, in which spin-dependent scattering and tunnelling are being successfully applied, as already mentioned in reading heads of high-density hard-discs and in magnetic random access memories (MRAM). However, in the context of spin electronics particularly interesting are ferromagnetic semiconductors, which combine complementary functionalities of ferromagnetic and semiconductor material systems. One of the relevant questions is to what extend the powerful methods developed to control the carrier concentration and spin polarization in semiconductor quantum structures could serve to tailor the magnitude and orientation of magnetization produced by the spins localized on the magnetic ions.

Another important issue concerns the elaboration of methods of injecting and transporting spin currents. In addition of consisting the important ingredient of field sensors and magnetic transistors, spin injection can serve as a tool for fast modulation of light polarization in semiconductors lasers.

Since the fabrication of quantum structures is most mature in the case of III-V semiconducting compounds, the milestone discovery was the detection of the carrier-induced ferromagnetism in $In_{1-x}Mn_xAs$ and $Ga_{1-x}Mn_xAs$ by IBM [20] and Tohoku University [21] groups, respectively. While the divalent Mn introduce both spins and holes in the III-V materials, the magnetic ion and carrier concentrations can be varied independently in II-VI materials, like in the case of IV-VI materials in which the hole-controlled ferromagnetism was put into the evidence in Warsaw already in 80s [22].

A systematic experimental and theoretical study of the carrier-induced ferromagnetism in II-VI semimagnetic semiconductors has been undertaken by the Grenoble-Warsaw collaboration [23]. In agreement with the theoretical model proposed by the team for various dimensionality II-VI semimagnetic semiconductors [23], the ferromagnetic order has been observed above 1 K in two dimensional modulation-doped p-type $Cd_{1-x}Mn_xTe/Cd_{1-y-z}Mg_yZn_zTe:N$ heterostructures [24,25]. The obtained results lead to suggest [24] that it will be possible to control magnetic properties by methods elaborated earlier to tune the carrier concentration in quantum structures, such as voltage biasing and light illumination. More recently, epitaxial layers of $Zn_{1-x}Mn_xTe:N$ with the hole concentration above $10^{20}$ cm$^{-3}$ have been obtained. By means of transport and magnetic measurements, the ferromagnetism was found in this three dimensional system [26,27], corroborated the theoretical predictions mentioned above [23]. Ferromagnetic correlation has been detected also in $Be_{1-x}Mn_xTe:N$ [28] as well as in bulk crystal of $Zn_{1-x}Mn_xTe:P$ [27,28]. At the same time, in agreement with the theoretical expectations [23], no ferromagnetism has been detected above 1 K in n-type films of $Zn_{1-x}Mn_xO:Al$ [27]. The stronger ferromagnetism in p-type materials comparing n-type compounds stems from a large magnitude of the hole density of states and a strong spin-dependent hybridization between the valence band p-like states and the Mn d orbitals. These two effects conspire to make the hole-mediated ferromagnetic interactions strong enough to overcome the antiferromagnetic super-exchange, specific to intrinsic DMS.

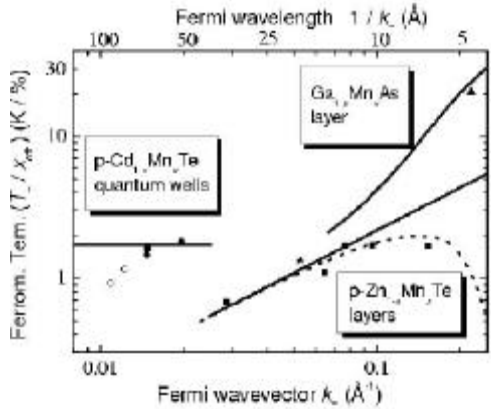

**Fig. 1.** Experimental (symbols) and calculated (lines) ferromagnetic temperatures (the sum of the Curie and antiferromagnetic temperatures) normalized by the effective Mn content versus the wave vector at the Fermi level for $Ga_{1-x}Mn_xAs$ (triangle) $Zn_{1-x}Mn_xTe:N$ (squares), $Zn_{1-x}Mn_xTe:P$ (star), and quantum well of p-$Cd_{1-x}Mn_xTe$ (circles). Solid lines: theory for the 3D and 2D cases; dotted line: theory taking into account the effect of antiferromagnetic interactions on statistical distribution of unpaired Mn spins (after [25-27,29,30]).

Theoretical studies undertaken simultaneously by the Warsaw-Tohoku collaboration lead to the elaboration of a theoretical model of thermodynamic, magneotoelastic and optical properties of III-V and II-II ferromagnetic semiconductors [26,29,30]. The model takes into account the *kp* and spin-orbit interactions, biaxial strain and confinement, it applies for both zinc-blende and wurzite compounds. It has been demonstrated that this model describes, with no adjustable parameters, the Curie temperature, the dependence of magnetization on temperature and the magnetic field, and the magnetic anisotropy energies in p-$Ga_{1-x}Mn_xAs$ [29,30] as well as the magnetic circular dichroism [30] that has been intensively studied in University of Warsaw [31]. As shown in Fig. 1, good agreement between experimental and theoretical results has been obtained also in the case of $Cd_{1-x}Mn_xTe/Cd_{1-y-z}Mg_yZn_zTe:N$ [25] and p-$Zn_{1-x}Mn_xTe$ [26]. The theoretical and numerical analysis [23-26,30,31] made it possible to describe quantitatively the dominant mechanism accounting for the ferromagnetism of the studied systems as well as for the demonstration of the role played by the space dimensionality, spin-orbit interaction, and interband polarization as well as for the identification of main differences between properties of III-V and II-VI compounds. Furthermore, theory of magnetic domains in III-V materials has been developed in collaboration with The University of Texas in Austin providing, among other things, the width of domains walls as well as the critical film size corresponding to the transition to a single domain structure [32]. Theoretical studies [29,30,32] have demonstrated that the spin-orbit coupling in the valence band, not at the magnetic ion, accounts for the magnetoelastic properties and magnetic anisotropy in the systems in question.

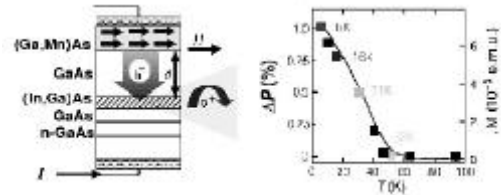

**Fig. 2.** The p-i-n electroluminescence diode with the (Ga,Mn)As p-type region, which emits the circularly polarized light in the absence of the external magnetic field. The degree of polarization (points, left scale) depends on the temperature in the same way as the spontaneous magnetization (solid line, right scale) (after [34]).

Another important progress that has recently been accomplished concerns the development of new tools enabling the spin manipulation. The tailoring of domain structures and magnetic anisotropy by confinement [24,25] and by strain engineering [30,32] – that is by employing substrates with appropriately adjusted lattice constants -- has been experimentally demonstrated. The recent optical detection of electrical spin injection in semiconductor p-i-n diodes using epitaxial structures with either a spin filter [33] or spin emitter [34] of diluted magnetic semiconductors has attracted a great deal of attention. Figure 2 presents experimental results concerning the injection of carriers to nonmagnetic quantum well of (In,Ga)As from p-type ferromagnetic (Ga,Mn)As and GaAs [34]. As shown, circular light polarization, which contains information on the degree of carrier spin polarization, is non-zero even in the absence of an external magnetic field, provided that the temperature is below the Curie point. As already mentioned, the efficient spin injection constitutes the important step on the way to construct magnetic switching devices, such as spin transistors [13] and quantum gates [15-17].

Another important step is the demonstration of control of magnetization by an electric field [35,36] or light [24,25,36,37]. The results presented in Fig. 3 show how the change of the hole concentration imposed by the gate bias generates a reversible and isothermal crossover between ferromagnetic and paramagnetic phases [35]. A similar effect has also been observed in a p-i-n diode containing (Cd,Mn)Te quantum well [36]. Interestingly, it has been also demonstrated that the light enhances the ferromagnetism in p-i-n diodes but destroys it in p-i-p structures, a behavior explained taking into account the distribution of the in-built electric field, and thus the distribution of the photocarriers, in these systems [36]. Hence, the magnetization of ferromagnetic semiconductor heterostructures can be alternated not only by the magnetic field or heat pulses, as in the existing magnetic memory devices, but also by a bias voltage or photon flux. Such a control opens an attractive perspective for integrating permanent memories with microprocessors. Indeed, there is a ground to hope that spintronics based on ferromagnetic semiconductors will lead to advances in sensors as well as in classical and, perhaps, quantum information devices.

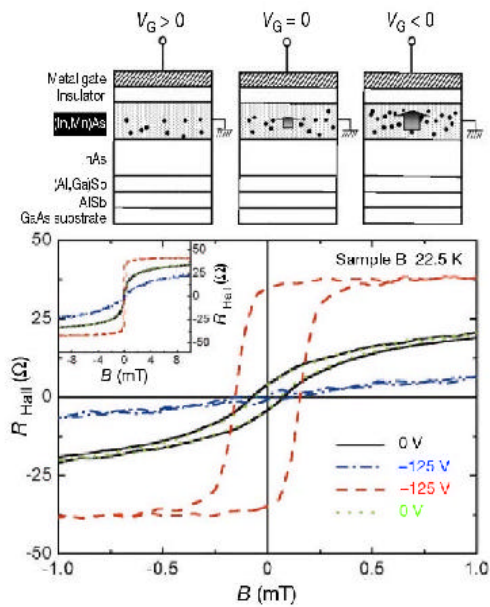

**Fig. 3.** Field effect transistor with the p-type (In,Mn)As channel. Depending on the magnitude of the gate voltage, and thus on value of the hole concentration, the Hall resistance (which is proportional to magnetization because of the anomalous Hall effect) shows either paramagnetic or ferromagnetic characteristics at constant temperature of 22.5 K. Inset shows the behaviour of the Hall resistance over a wider field range (after [35]).

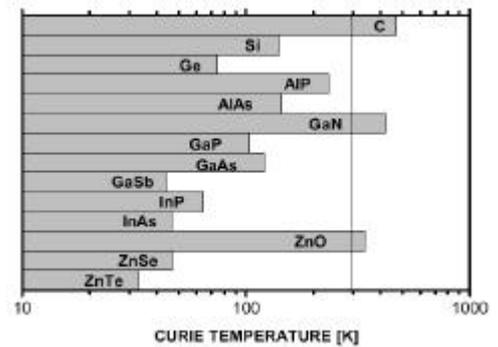

**Fig. 4.** Computed values of the Curie temperature for various p-type semiconductors containing 5% of Mn per cation (2.5% per atom) and $3.5 \times 10^{20}$ holes per cm$^3$ (after [29,30]).

One has to emphasize that the results described above have been obtained for magnetic semiconductors in which the highest value of Curie temperature does not exceed 110 K [21]. For this reason, the theoretical work has been undertaken [29] aiming at evaluation the magnitudes of expected Curie temperatures in various III-V and II-VI compounds as well as in elemental group IV semiconductors [29,30]. The results shown in Fig. 4 demonstrate that in semiconductors consisting of light elements, the critical point may exceed the room temperature [29,30]. These predictions have encouraged many groups to synthesize such materials as GaN and ZnO containing Mn or other transition metals as well as to search of ferromagnetic elemental semiconductors. Before presenting promising experimental results we have to caution the readers that it some cases the observed ferromagnetism might have been resulted from ferromagnetic or ferrimagnetic inclusions or precipitates.

Among interesting results worth quoting is the discovery of ferromagnetism in (Ge,Mn) [38] as well as the observation of the ferromagnetic order above 300 K in (Cd,Mn,Ge)P$_2$ [39], (Zn,Co,V)O [40], (Ti,Co)O$_2$ [41], and (Ga,Mn)N [42]. In the case of the ferromagnetic (Ga,Mn)N [42,43] the record high extrapolated Curie temperature $T_C$ is 940 K for the Mn content of the order of 9% [42], which suggests that the relevant ferromagnetic interactions are stronger than in Co, which has the highest known Curie temperature, $T_C = 1400$ K. This demonstrates the existence of unusually strong coupling between magnetic moments, which can be assigned to strong p-d hybridization [43]. A particularly interesting family constitutes materials, such as (Ca,La)B$_6$ [44] and polimerized C$_{60}$ [45], which do not contain any magnetic elements but nevertheless exhibit spontaneous magnetization persisting up to temperatures well above 300 K.

There is no doubt that because of prospects the spintronics offers for basic and applied physics, searches for high temperature ferromagnetic semiconductors have evolved into the important branch of materials science.

**Acknowledgments**

The works in the field of ferromagnetic semiconductors reported here have been performed in Warsaw in collaboration with Grenoble University (the group of J. Cibert), Tohoku University in Sendai (the group of H. Ohno), and The Unversity of Texas (the group of A.H. MacDonald). The research was supported by Foundation for Polish Science, by State Committee for Scientific Research (Grant No.~2-P03B-02417 and Polonium Program) as well as by FENIKS project (EC: G5RD-CT-2001-00535).